\documentclass[sigconf,authorversion,nonacm]{acmart}

\usepackage{graphicx}
\usepackage{subcaption}
\usepackage{algorithmic}
\usepackage{algorithm}
\usepackage{dsfont}
\usepackage{balance} 
\usepackage{adjustbox}

\begin{document}

\title{Fast generation of simple directed social network graphs with reciprocal edges and high clustering}

\author{Christoph Schweimer}
\email{cschweimer@know-center.at}
\affiliation{%
  \institution{Know-Center GmbH}
  \city{Graz}
  \country{Austria}
}

\renewcommand{\shortauthors}{Schweimer}

\begin{abstract}
Online social networks have emerged as useful tools to communicate or share information and news on a daily basis. One of the most popular networks is Twitter, where users connect to each other via directed follower relationships. Twitter follower graphs have been studied and described with various topological features. Collecting Twitter data, especially crawling the followers of users, is a tedious and time-consuming process and the data needs to be treated carefully due to its sensitive nature, containing personal user information. We therefore aim at the fast generation of directed social network graphs with reciprocal edges and high clustering.

Our proposed method is based on a previously developed model, but relies on less hyperparameters and has a significantly lower runtime. Results show that our method does not only replicate the crawled directed Twitter graphs well regarding several topological features and the application of an epidemics spreading process, but that it is also highly scalable which allows the fast creation of bigger graphs that exhibit similar properties as real-world networks.
\end{abstract}


\maketitle

\section{Introduction}

Online social networks, e.g., Facebook, Instagram or Twitter are frequently used media to read news and share information in a quick and condensed way. The user graphs have been studied and are often described at the hand of various topological graph features (e.g., average shortest path length, diameter, clustering coefficient). Getting access to social networks data is a tedious and resource-intense process which might not be affordable for some research groups. The data of the ever evolving social network graphs is also only a snapshot and needs to be treated with a lot of care due to the personal user information that is linked to it. We therefore saw the need for a fast method to generate big directed social network graphs with topological features similar to real-world network graphs. This work is building on a previous approach \cite{Schweimer2022} to create directed social network graphs for information spreading and therefore uses parts of the method to generate the graphs. We sample correlated reciprocal, in- and out-degrees for each node and connect the nodes according to the sampled degrees without generating self-loops (node connected to itself) or parallel edges (two edges between two nodes in the same direction). New neighbors of a node are connected directly to achieve high clustering, making the method highly efficient.

We used our method to replicate 14 crawled Twitter follower subgraphs of wide-ranging sizes between 500 and 50,000 users. Results show that our method generates graphs that are similar to their real-world counterparts w.r.t.\ various topological features and when simulating an epidemics spreading process on them. We also created and analysed a graph with about twice as many nodes as the biggest crawled graph to show the scalability of our method.

The rest of the paper is structured as follows. In Section \ref{RELATED_WORK} we review graph generation methods. In Section \ref{CRAWLED_NETWORKS} we briefly describe the crawled Twitter subgraphs and in Section \ref{METHOD} we present the graph generation method. In Section \ref{RESULTS} we go over the results and Section \ref{DISCUSSION} concludes the paper.

\section{Related Work}
\label{RELATED_WORK}

The generation of random graphs has been studied long before the emergence of online social networks. Classic approaches include amongst others the \textit{Erd\H{o}s-R\'enyi} model \cite{Erdos1959,Gilbert1959}, the \textit{Barab\'asi-Albert} model \cite{Barabasi1999} and the \textit{Watts-Strogatz} model \cite{Watts1998}. \cite{Bonifati2020} summarise a variety of state of the art graph generation approaches.
Various approaches connect nodes based on sequences of node degrees. In the \textit{Chung-Lu} model \cite{Chung2002,Durak2013}, two nodes connect to each other in a probabilistic way proportional to their degrees to create an edge, whereas in the \textit{Configuration} model \cite{Bender1978, Bollobas1980, Britton2006, Chen2013}, the nodes develop stubs, depending on their degree, and two of them connect to create an edge until all stubs are used.

Many social network graphs, like Twitter, contain a lot of reciprocal edges \cite{Durak2013, Kwak2010} and a high amount of clustering \cite{Myers2014,Ahn2007,Mislove2007}, which are only partially being considered in the random graph generation approaches above. \cite{Newman2009} introduces a method to create undirected graphs with high clustering, and \cite{Durak2013} propose a null model that considers reciprocal and directed edges. \cite{Schweimer2022} developed a model, which creates simple directed social network graphs including reciprocal edges for information spreading. The idea is to sample correlated node degrees from $\chi^2$-distributions for the reciprocal, in- and out-degree of each node, create reciprocal and directed edges separately with a Chung-Lu model like approach \cite{Durak2013} and apply an edge rewiring procedure \cite{Bansal2009} on selected nodes to increase the clustering coefficient to a real-world level. The results indicate that the crawled Twitter follower networks, that their work is based on, are replicated well w.r.t.\ several topological features and algorithmic properties. The results also show that there is a discrepancy in the rank correlations between the degrees between the crawled and created graphs and that the runtime exceeds multiple days for a graph with approximately 50,000 nodes. We thus developed a fast model to generate simple directed social network graphs with reciprocal edges and high clustering that has a short runtime and preserves the rank correlations between the degrees more accurately. 

Our method builds on the approach presented in \cite{Schweimer2022} by sampling correlated reciprocal, in- and out-degrees, but instead of connecting nodes with the principle of the Chung-Lu model \cite{Chung2002}, we utilise the principle of the Configuration model \cite{Bender1978, Bollobas1980, Britton2006, Chen2013} and instead of applying an edge rewiring procedure \cite{Bansal2009} after sampling all edges to achieve high clustering, we establish connections between new neighbors of a node directly, thus decreasing the runtime significantly.
Results show that our new method creates graphs with high clustering and with similar topological features and algorithmic properties as real-world social network graphs. We can also efficiently generate bigger graphs with a high number of nodes and edges.

\section{Crawled Twitter Follower Graphs}
\label{CRAWLED_NETWORKS}

Our work is based on \cite{Schweimer2022} and our goal is to replicate the 14 crawled Twitter follower subgraphs (approx. 500 - 50,000 users) described therein. The graphs $G = (V,E)$ are directed, the nodes $v \in V$ represent the users and the edges $e \in E$ represent the directed follower relationships between the users. If two users follow each other, we merge the two directed edges with opposite direction into one reciprocal edge.
We computed several topological features to describe the graphs, e.g., the largest connected components, the density and the average clustering coefficient. Additionally, we also computed the rank correlations (Spearman's $\rho$) between the degrees. The numbers are listed in the first line of each cell in Table \ref{Results_all} and features computed in the largest weakly connected component are marked with an asterisk (*).

\section{Graph Generation Method}
\label{METHOD}

We generate directed social network graphs with reciprocal edges and high clustering with the same number of nodes as the crawled Twitter follower subgraphs\footnote{For the code see \url{https://github.com/Buters147/Social_Network_Graph_Generator}} and aim at replicating them w.r.t.\ their topological features and improving the runtime of the method in \cite{Schweimer2022}.
We therefore adapt that approach by connecting nodes with the principle of the Configuration model, instead of the Chung-Lu model and not applying an edge rewiring procedure.
 
\subsection{Node Degree Sampling}
\label{NODE_DEGREE_GENERATION}

For each node in a graph with $n = \lvert V \rvert$ nodes, we sample correlated reciprocal, in- and out-degrees from fitted $\chi^2$-distributions as described in \cite{Schweimer2022} and generate edges between nodes with the principle of the Configuration model. To connect all stubs, the reciprocal degree sum must be an even number and the in-degree sum must be equal to the out-degree sum \cite{Chen2013}. For the first issue, we look at the reciprocal degree sum and if it is an odd number, we increase the reciprocal degree of the node with the highest reciprocal degree by one.
For the second issue, we repeat the node degree sampling until the difference $\tau$ between the sampled in-degree and out-degree sum is less than a predefined threshold (we are using $10\%$ of the number of nodes in the graph, i.e., $\tau < 0.1 \cdot n$, to not affect the rank correlations too much). To equalise the two sums, we randomly choose $\tau$ different nodes without replacement and increase their node degree that contributes to the lower degree sum by one. 

This sampling and equalising procedure returns the generated reciprocal, in- and out-degree sequences $\mathbb{R}, \mathbb{I}, \mathbb{O}$ of length $n$ where $\sum_i \mathbb{R}_i$ is an even number and $\sum_i \mathbb{I}_i = \sum_i \mathbb{O}_i$. 

\subsection{Edge Sampling}
\label{CONNECTING}

\begin{figure}
\small
\centering
\includegraphics[width=1\linewidth]{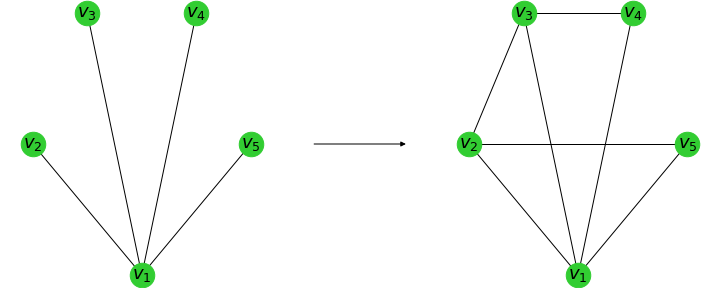}
\caption{Two-step procedure for node $v_1$ with reciprocal edges, and $v_2, v_3, v_4, v_5$ being new first degree neighbors}
\label{Two-step}
\end{figure}

We initialise the null graph $G = (V,E)$ with $n = \lvert V \rvert$ nodes $(V = \{v_1, \ldots, v_n\})$ and the reciprocal, in- and out-degree sequences $\mathbb{R}, \mathbb{I}, \mathbb{O}$. To generate edges, we use the principle of the Configuration model by connecting stubs of nodes. 

We start by generating reciprocal edges without self-loops (node connected to itself) and parallel edges (two edges between two nodes in the same direction) in an iterative two-step procedure. In the first step of each iteration, we search for the node with the maximum reciprocal degree $\max(\mathbb{R})$, connect as many stubs of that node to one stub of other nodes (i.e., no self-loops) without sampling parallel edges and reduce their value in $\mathbb{R}$ accordingly. This step is unweighted, i.e., the probability for the node with the highest degree to connect to another node is independent of the sampled degree of the other nodes. In the second step, we look at the nodes that established a new connection in the current iteration (we call them new first degree neighbors), randomly connect them to each other with a reciprocal edge, based on their degree, but without generating parallel edges (i.e., if there is already an edge between two nodes, we cannot connect them again) and reduce their value in $\mathbb{R}$ accordingly\footnote{Connecting new first degree neighbors not only with reciprocal edges, but also with directed edges lead to increased values for the CC, exceeding 0.6, which is unrealistic for social network graphs.}. This two-step procedure is repeated until no further reciprocal edges can be generated, leading to an undirected graph. The sampled edges are implemented as two directed edges with opposite direction. The two-step procedure for node $v_1$ with reciprocal edges is depicted in Fig. \ref{Two-step}, where $v_2, v_3, v_4, v_5$ emerged as new first degree neighbors in the first step and randomly established connections to each other in the second step.

Generating directed edges follows the same iterative two-step procedure. We search for the node with the highest out-degree (note that we could also search for the node with the highest in-degree) and connect as many outgoing stubs of that node to incoming stubs of other nodes without sampling parallel edges (i.e., we cannot sample edges that have already been sampled). Next, we randomly connect new first degree neighbors via directed edges. Node degrees ($\mathbb{O}$ and $\mathbb{I}$) are reduced accordingly in each step. Note that edges in both directions can be created, leading to additional reciprocal edges. This two-step procedure is again repeated until no further directed edges can be generated.

\section{Results}
\label{RESULTS}

\subsection{Topological Features}

\begin{table*}
\caption{Topological features of 14 crawled graphs (line 1), the corresponding new graphs (line 2) and the old graphs (line 3)} 
\begin{adjustbox}{width=\textwidth,center}
 \begin{tabular}{l c c c c c c c c c c c c c c} 
\toprule
  & $G_1$ & $G_2$ & $G_3$ & $G_4$ & $G_5$ & $G_6$ & $G_7$ & $G_8$ & $G_9$ & $G_{10}$ & $G_{11}$ & $G_{12}$ & $G_{13}$ & $G_{14}$\\ 
 \midrule
 Nodes & 11,015 & 21,291 & 50,133 & 459 & 3,580 & 8,277 & 21,464 & 13,646 & 2,013 & 15,299 & 6,003 & 2,464 & 1,239 & 2,932 \\ 
 \midrule
 
  & 377,457 & 2,570,452 & 4,832,226 & 5,435 & 54,735 & 791,905 & 530,302 & 517,916 & 18,781 & 692,534 & 223,175 & 43,572 & 30,285 & 44,261\\ 
  
 Edges & 388,626 & 2,391,103 &	4,962,837 &	5,349 & 51,708 & 750,358 & 547,170 &	525,738 & 17,726 &	694,396 & 213,554  &	41,486 & 30,922 & 41,932\\
 
  & 381,627 & 2,527,541 & 5,049,608 & 5,499 & 57,249 & 799,792 & 541,986 & 529,675 & 17,042 & 692,364 & 218,306 & 41,980 & 28,819 & 44,627\\ 
 \midrule
 
  & 0.0031 & 0.0057 & 0.0019 & 0.0259 & 0.0043 & 0.0116 & 0.0012 & 0.0028 & 0.0046 & 0.0030 & 0.0062 & 0.0072 & 0.0197 & 0.0052\\
  
 Density & 0.0032 &	0.0053 &	0.0020 & 0.0254	 & 0.0040	 & 0.0110 &	0.0012 &	0.0028 & 0.0044	 &	0.0030 & 0.0059 & 0.0068	 &	0.0202 & 0.0049\\
 
  & 0.0031 & 0.0056 & 0.0020 & 0.0262 & 0.0045 & 0.0117 & 0.0012 & 0.0028 & 0.0042 & 0.0030 & 0.0061 & 0.0069 & 0.0188 & 0.0052\\ 
  \midrule
 
  & 9,347 & 19,237 & 43,461 & 361 & 2,314 & 7,352 & 12,793 & 9,977 & 1,154 & 13,001 & 5,286 & 2,100 & 1,094 & 1,917\\ 
  
 LSCC & 9,215 &	19,927 &	44,296 & 424	 & 2,443	 &  7,315 &	13,663 &	10,901 & 1,220	 &	13,263 & 5,489  & 2,151	 &	1,205 &	2,002\\
 
  & 8,182 & 19,296 & 41,805 & 390 & 1,999 & 7,105 & 11,411 & 9,639 & 990 & 12,173 & 5,009 & 1,882 & 1,141 & 1,686\\
 \midrule
 
  & 10,931 & 21,281 & 49,999 & 452 & 3,570 & 8,272 & 21,418 & 13,631 & 1,912 & 15,288 & 5,964 & 2,443 & 1,224 & 2,797\\ 
  
 LWCC & 10,561 &	21,156 &	48,948 & 454	 & 3,510	 & 8,172 &	21,002 &	13,305 & 1,745	 &	14,946 & 5,847  & 2,408	 & 1,221	 & 2,680	\\
 
  & 10,166 & 21,148 & 48,607 & 446 & 3,354 & 8,203 & 19,946 & 13,022 & 1,609 & 14,574 & 5,805 & 2,290 & 1,222 & 2,493\\
 \midrule

  & 0.0032 & 0.0057 & 0.0019 & 0.0266 & 0.0043 & 0.0116 & 0.0011& 0.0028 & 0.0051 & 0.0030 & 0.0063 & 0.0073 & 0.0202 & 0.0056\\ 
  
 Density* & 0.0035 &	0.0053 &	0.0021 & 0.0260	 & 0.0042	 &  0.0112 &	0.0012 &	0.0030 & 0.0058	 &	0.0031 & 0.0062 &	0.0072 &	0.0208 & 0.0058	\\
 
  & 0.0037 & 0.0057 & 0.0021 & 0.0277 & 0.0051 & 0.0119 & 0.0014 & 0.0031 & 0.0066 & 0.0033 & 0.0065 & 0.0080 & 0.0193 & 0.0072\\
  \midrule
 
  & 2.97 & 2.69 & 2.75 & 2.40 & 2.34 & 2.52 & 2.49 & 2.69 & 2.65 & 2.65 & 2.70 & 3.04 & 2.68 & 2.41\\
  
 ASPL* & 2.63 &	2.50 &	2.61 & 2.52	 & 2.25	 & 2.27 &	2.15 &	2.49 & 2.35	 &	2.65 & 2.69  & 2.72	 & 2.61	 & 2.36	 \\
 
  & 2.59 & 2.70 & 2.79 & 2.48 & 1.99 & 2.33 & 2.00 & 2.38 & 2.11 & 2.64 & 2.68 & 2.55 & 2.59 & 2.18\\
 \midrule
 
  & 11 & 9 & 10 & 8 & 11 & 7 & 10 & 10 & 11 & 9 & 9 & 10 & 7 & 11\\
  
 Diameter* & 4 & 4 & 4 & 4 & 4 & 4 & 5 & 4 & 5  & 4 & 4  & 5  & 4 & 5 \\  
 
  & 7 & 6 & 7 & 5 & 6 & 5 & 7 & 7 & 7 & 7 & 7 & 7 & 5 & 7\\
 \midrule
 
  & 0.201 & 0.285 & 0.198 & 0.354 & 0.277 & 0.339 & 0.266 & 0.244 & 0.260 & 0.319 & 0.234 & 0.321 & 0.303 & 0.310\\
  
 Average CC* & 0.315 &	0.352 &	0.320 &	0.365 &	0.300 & 0.389 &	0.304 &	0.309 & 0.236	 &	0.346 & 0.321  &	0.300 &	0.369 &	0.292\\
 
  & 0.228 & 0.291 & 0.222 & 0.300 & 0.187  & 0.380 & 0.150 & 0.215 & 0.176 & 0.240 & 0.240 & 0.251 & 0.312 & 0.210\\
\midrule
 
 & 0.540 & 0.630 & 0.616 & 0.465 & 0.407 & 0.619 & 0.420 & 0.489 & 0.262 & 0.586 & 0.658 & 0.592 & 0.656 & 0.419\\
 
 $\rho_1$  & 0.522 &	0.623 &	0.605 &	0.456 &	0.383 & 0.602 &	0.396 &	0.469 &	0.233 &	0.574 & 0.639 &	0.589 &	0.641 &	0.394\\
 
  & 0.470 & 0.604 & 0.579 & 0.259 & 0.319 & 0.578 & 0.345 & 0.425 & 0.167 & 0.523 & 0.578 & 0.457 & 0.544 & 0.333\\
 \midrule
 
  & 0.612 & 0.697 & 0.753 & 0.606 & 0.547 & 0.707 & 0.605 & 0.664 & 0.253 & 0.584 & 0.564 & 0.577 & 0.611 & 0.315\\
  
 $\rho_2$  & 0.562 & 0.692 &	0.741 &	0.555 &	0.443 & 0.704 &	0.511 &	0.635 &	0.221 &	0.559 & 0.550 &	0.540 &	0.583 &	0.303\\
 
  & 0.528 & 0.660 & 0.696 & 0.419 & 0.411 & 0.676 & 0.474 & 0.564 & 0.192 & 0.503 & 0.476 & 0.486 & 0.496 & 0.245\\
 \midrule
 
  & 0.284 & 0.395 & 0.397 & 0.240 & 0.250 & 0.424 & 0.349 & 0.297 & -0.001 & 0.400 & 0.242 & 0.258 & 0.267 & 0.148\\
  
 $\rho_3$  & 0.254 &	0.401 &	0.411 &	0.224 &	0.200 & 0.419 &	0.318 &	0.309 &	-0.016 &	0.401 & 0.268 &	0.224 &	0.288 &	0.133\\
 
  & 0.234 & 0.369 & 0.371 & 0.140 & 0.149 & 0.370 & 0.266 & 0.238 & -0.035 & 0.337 & 0.182 & 0.192 & 0.170 & 0.135\\
   \midrule
 
  Runtime & 15min &	70min &	4h & 10s	 &	3min & 30min &	45min &	23min &	30s &	22min & 9min &	45s &	30s & 1min	\\
  
   & 1h & 31h & 110h & 1min & 3min & 9h & 2h & 2h & 1min & 2.5h & 1h & 5min & 2min & 4min\\
\bottomrule
\end{tabular}
\end{adjustbox}
\label{Results_all}
\end{table*}

The topological features of the crawled graphs (line 1 in each cell), the created graphs with the proposed method (line 2 in each cell), which we call new method, and the created graphs with the method in \cite{Schweimer2022} (line 3 in each cell), which we call old method, are listed in Table \ref{Results_all} for comparison. Note that for the runtime we only have results for the new method (line 1) and the old method (line 2).

The numbers in Table \ref{Results_all} show that our new approach creates similar graphs compared to the crawled graphs. The sizes of the largest strongly connected components (LSCC) and the largest weakly connected components (LWCC) are replicated well and we observe that the LSCC tends to be bigger, whereas the LWCC tends to only be a bit smaller in the new graphs. In the graphs created with the old method, we see a higher discrepancy regarding the sizes of the largest connected components, especially the LWCC. With the new method, the sampled node degrees are preserved more accurately by using the principle of the Configuration model. Since we avoid sampling self-loops and parallel edges, there are still nodes, where the sampled node degrees are not met exactly, leading to nodes having a smaller degree than sampled and to nodes having a degree of zero.

The average shortest path length (ASPL) is in a similar range in the newly created graphs, tending to be smaller, and they have a much smaller diameter (either 4 or 5). This is caused by directly creating edges between the new first degree neighbors (second part of the two-step procedure). By adding nodes on the periphery of a graph, we could artificially increase the diameter.

The average CC of the graphs with the new method is in a real-world range (between 0.24 and 0.39), which is slightly higher than in the crawled graphs (0.20 to 0.35). We see that, with the exception of one Graph ($G_9$), the CC is close to or exceeding 0.30. This high CC, which is a common trait in social network graphs \cite{Myers2014,Ahn2007,Mislove2007}, is again caused by the second part of the two-step procedure to create edges between new first degree neighbors, as connecting two neighbors of a node increases the CC of these three nodes.

We see that the graphs with the new method are much more similar to the crawled graphs regarding the rank correlations $\rho_1$ (reciprocal, in), $\rho_2$ (reciprocal, out) and $\rho_3$ (in, out) between the degrees. In the graphs with the old method, the nodes were connected with the principle of the Chung-Lu model, which relies on probabilistic edge sampling, often leading to nodes having a higher or lower degree than sampled and also to several nodes having a degree of zero. By using the principle of the Configuration model, the sampled node degrees cannot be exceeded and are preserved more accurately. Since we avoid sampling self-loops and parallel edges, not all stubs are being connected for some nodes, causing a small discrepancy in the rank correlations.

\subsection{Algorithmic Properties}

\begin{table*}
\caption{Algorithmic properties (SIR simulation) of crawled graphs (line 1), the corresponding new graphs (line 2) and the old graphs (line 3)} 
\begin{adjustbox}{width=\textwidth,center}
 \begin{tabular}{l c c c c c c c c c c c c c c c c} 
\toprule
 & & $G_1$ & $G_2$ & $G_3$ & $G_4$ & $G_5$ & $G_6$ & $G_7$ & $G_8$ & $G_9$ & $G_{10}$ & $G_{11}$ & $G_{12}$ & $G_{13}$ & $G_{14}$ & $100k$\\ 
  \midrule
 Nodes & & 11,015 & 21,291 & 50,133 & 459 & 3,580 & 8,277 & 21,464 & 13,646 & 2,013 & 15,299 & 6,003 & 2,464 & 1,239 & 2,932 & 100,000\\ 
 \midrule
  & & 0.121 & 0.415 & 0.365 & 0.057 & 0.062 & 0.382 & 0.098 & 0.135 & 0.059 & 0.190 & 0.128 & 0.068 & 0.074 & 0.070 & - \\ 
 $p=0.01$ & $\%$ & 0.148 & 0.442 & 0.393 & 0.058 & 0.061 & 0.386 & 0.094 & 0.158 & 0.057 & 0.212 & 0.124 & 0.066 & 0.072 & 0.065 & 0.617 \\ 
  & & 0.152 & 0.447 & 0.388 & 0.057 & 0.064 & 0.397 & 0.099 & 0.163 & 0.057 & 0.212 & 0.118 & 0.069 & 0.070 & 0.071 & - \\
  \midrule
    & & 13.18 & 12.00 & 13.15 & 3.31 & 5.88 & 11.57 & 13.97 & 15.09 & 5.21 & 13.74 & 13.66 & 7.07 & 6.77 & 7.62 & - \\ 
  & steps & 15.99 & 11.34 & 11.69 & 3.43 & 5.55 & 11.52 & 14.27 & 16.34 & 4.46 & 15.78 & 14.18 & 6.41 & 6.36 & 6.49 & 8.66 \\
  & & 15.85 & 11.10 & 11.71 & 3.37 & 5.94 & 11.65 & 14.65 & 15.45 & 4.54 & 15.19 & 13.26 & 7.22 & 6.27 & 8.02 & - \\
  \midrule
  \midrule
  & & 0.460 & 0.760 & 0.734 & 0.219 & 0.330 & 0.731 & 0.389 & 0.515 & 0.178 & 0.562 & 0.529 & 0.333 & 0.437 & 0.273 & - \\ 
 $p=0.05$ & $\%$ & 0.507 & 0.752 & 0.686 & 0.197 & 0.336 & 0.708 & 0.479 & 0.554 & 0.193 & 0.568 & 0.545 & 0.356 & 0.494 & 0.291 & 0.828 \\ 
  & & 0.503 & 0.762 & 0.689 & 0.220 & 0.356 & 0.742 & 0.480 & 0.555 & 0.189 & 0.576 & 0.535 & 0.361 & 0.456 & 0.299 & - \\
  \midrule
   & & 12.43 & 8.65 & 9.77 & 9.85 & 11.22 & 9.18 & 11.20 & 10.51 & 9.71 & 9.59 & 9.50 & 10.65 & 9.63 & 11.01 & - \\ 
  & steps & 8.31 & 6.81 & 7.18 & 9.50 & 11.84 & 6.64 & 9.42 & 8.41 & 11.15 & 7.87 & 8.28 & 10.65 & 9.69 & 9.39 & 6.32 \\ 
  & & 8.65 & 7.13 & 7.41 & 10.24 & 10.76 & 7.00 & 9.68 & 8.79 & 10.65 & 8.25 & 9.17 & 9.58 & 9.99 & 8.98 & - \\
  \midrule
  \midrule
  & & 0.638 & 0.857 & 0.837 & 0.496 & 0.530 & 0.838 & 0.564 & 0.676 & 0.282 & 0.700 & 0.685 & 0.510 & 0.600 & 0.410 & - \\ 
 $p=0.1$ & $\%$ & 0.633 & 0.837 & 0.778 & 0.533 & 0.536 & 0.800 & 0.633 & 0.687 & 0.331 & 0.693 & 0.686 & 0.519 & 0.669 & 0.426 & 0.886 \\ 
  & & 0.636 & 0.853 & 0.785 & 0.524 & 0.551 & 0.845 & 0.640 & 0.693 & 0.324 & 0.708 & 0.682 & 0.527 & 0.646 & 0.435 & - \\
  \midrule
  & & 10.65 & 8.27 & 9.20 & 9.38 & 9.22 & 8.15 & 10.12 & 9.80 & 9.27 & 8.84 & 8.72 & 9.34 & 8.53 & 9.95 & - \\ 
  & steps & 7.04 & 6.07 & 6.26 & 9.65 & 8.70 & 5.86 & 7.62 & 7.09 & 8.76 & 6.86 & 6.99 & 7.81 & 7.26 & 7.84 & 5.81 \\ 
  & & 7.53 & 6.48 & 7.00 & 9.81 & 8.68 & 6.23 & 8.23 & 7.55 & 8.69 & 7.32 & 7.79 & 8.09 & 7.91 & 8.13 & - \\
\bottomrule
\end{tabular}
\end{adjustbox}
\label{ALGO_results}
\end{table*}

Social networks are used to circulate messages and information and the graphs representing them can be analysed by simulating how diseases or epidemics spread in them. We therefore compare the crawled and created graphs by simulating a version of the \textit{Susceptible-Infected-Recovered} (SIR) model \cite{Het00,Kiss2017} on the largest weakly connected component. At the start of a simulation a certain percentage (we chose $5\%$) of nodes in the graph is infected with a disease. At discrete time steps, infected nodes transmit the disease to their susceptible outgoing neighbors with transmission probability $p$ (we are using $p \in \{0.01, 0.05, 0.1\}$ in our experiments), and the nodes that were infected become recovered. A recovered node can then not be infected again and the simulation ends when there are no more infected nodes (i.e., the disease dies out).
We are using the EoN (Epidemics on Networks) Python module \cite{Miller2019} for the simulation of the discrete SIR model. We performed 100 simulations on the LWCC of each graph and used the fraction of recovered nodes (the number of nodes in the LWCC differs between crawled and created graphs) as well as the number of time steps until the end of each simulation as performance measures. The averaged results over the 100 simulations for the crawled graphs (line 1 of each cell), the graphs created with the new method (line 2 of each cell) and the graphs created with the old method (line 3 in each cell) are listed in Table \ref{ALGO_results} for comparison. 

We see that the fraction of recovered nodes tends to be slightly higher in the newly created graphs, but the difference is below 0.03 in most cases, especially for the lowest transmission probability $p=0.01$. The biggest difference can be seen in graph $G_7$ with 0.09 for $p=0.05$. The number of steps until the simulation ends (i.e., the disease dies out) is also in a similar range for the graphs. The graphs created with the new method tend to need slightly less steps, in particular for the higher transmission probabilities, which is likely caused by the higher clustering. The behavior in the graphs created with the old method and the new method is nearly identical in almost all cases.

\subsection{Runtime}

Experiments were carried out on 1 core of an Intel(R) Xeon(R) 6248 CPU @ 2.50GHz processor with 256GB RAM. The runtimes for the graph generation method are listed in the last cell of Table \ref{Results_all}. By comparing the times for the graphs $G_2$ and $G_7$ (approx. same number of nodes) and $G_7$ and $G_8$ (approx. same number of edges), we observe that the computation time depends on the number of nodes in the graph as well as the number of edges in the graph. 

One iteration of the two-step procedure to generate reciprocal edges consists of connecting the node with the highest reciprocal degree to other nodes and randomly connecting new first degree neighbors while checking if an edge can be generated. Searching for the node with the highest degree and connecting it to other nodes have a time complexity of $\mathcal{O}(n)$ respectively, while the second step depends on the maximum reciprocal degree and has a time complexity of $\mathcal{O}(\max(\deg(\mathbb{R}))^2)$. The degrees in $\mathbb{R}$ are reduced accordingly in each step and the runtime decreases in each new iteration. The theoretical time complexity is therefore $\mathcal{O}(n \cdot \max (\deg(\mathbb{R}))^2)$, but since the degrees in $\mathbb{R}$ are reduced in each iteration, the number of iterations is much lower than the number of nodes $n$ in the graphs. For directed edges, the procedure works accordingly. The method that our work is based on has a time complexity of $\mathcal{O}(n \cdot \max(\deg(V'))^2 \cdot \log \max(\deg(V')))$ \cite{Schweimer2022}, where $V'$ denotes the set of nodes that are considered for the edge rewiring procedure, which is higher than for the new method.

Looking at Table \ref{Results_all}, the runtimes of the new method are significantly shorter compared to the old method, mainly for big graphs where the reduction ranges from hours to several days. The creation of graph $G_3$ takes approx. 4h, whereas it took approx. 110h with the old method. This clearly shows the efficiency of our new method. 

\subsection{Bigger graph}

\begin{table}
\small
\caption{Topological features of a graph with 100,000 nodes} 
\centering 
 \begin{tabular}{l c c c c c c c c c c c c c c} 
\toprule
 Nodes & 100,000 \\ 
 \midrule
 Edges & 27,429,367\\
 \midrule
 Density & 0.0027\\
  \midrule
 LSCC & 97,303\\
 \midrule
 LWCC & 99,302\\
 \midrule
 Density* & 0.0028\\
  \midrule
 ASPL* & 2.61 \\
 \midrule
 Diameter* & 4\\  
 \midrule
 Average CC* & 0.373\\
 \midrule
  Runtime & 18h\\
\midrule
 $\rho_1$  & 0.598\\
 \midrule
 $\rho_2$  & 0.593\\
 \midrule
 $\rho_3$  & 0.502\\
 \bottomrule
\end{tabular}
\label{Results_100}
\end{table}

Besides replicating the crawled graphs mentioned in Section \ref{CRAWLED_NETWORKS}, we also generated a graph that contains 100,000 nodes to show the scalability of our method, with $\rho_1 = 0.6$, $\rho_2 = 0.6$, $\rho_3 = 0.5$  for the rank correlations between the degrees as input.

The topological features for this graph are listed in Table \ref{Results_100}. We see that the rank correlations $\rho_1, \rho_2, \rho_3$ are almost reproduced exactly which indicates that bigger graphs preserve the rank correlations better than smaller graphs. The graph has a high CC (0.373), but it is still on a real-world level and the diameter is again small (cf. Table \ref{Results_all}). The average shortest path length of the graph is 2.61, which is also on a realistic level comparing it to the other graphs (cf. Table \ref{Results_all}). Even though the computation time (approx. 18h) seems rather high, the result shows that our method is highly scalable. Gathering a real-world dataset of this size by crawling the followers of 100,000 users would take several weeks and creating a graph of that size with the method in \cite{Schweimer2022} would likely take several weeks as well. We simulated the SIR model on the graph, see last column of Table \ref{ALGO_results}, for completion.

\section{Conclusion}
\label{DISCUSSION}

We presented a method for the fast generation of simple directed social network graphs with reciprocal edges and high clustering. We adapt and improve a previously developed model \cite{Schweimer2022} by establishing connections between nodes according to their sampled degrees and directly connecting newly established neighbors to increase the clustering coefficient, thus avoiding the application of an edge rewiring procedure.

The results show that the generated graphs exhibit similar numbers for the topological features and algorithmic properties compared to the 14 crawled Twitter follower subgraphs. The generated graphs have a smaller diameter which is caused by the second part of the two-step procedure to sample edges and the average clustering coefficient in the largest weakly connected component tends to be a bit higher but is in a real-world range. The rank correlations are preserved more accurately compared to the method in \cite{Schweimer2022}, which comes from generating edges between nodes with the principle of the Configuration model, instead of the Chung-Lu model. A major improvement is the shorter runtime. Especially big graphs are generated much faster and we were able to decrease the runtime for the biggest graph from several days to approximately 4 hours. The created graphs also behave similar compared to the crawled graphs when applying an epidemics spreading algorithm (discrete SIR model) on them. In addition to replicating the crawled Twitter follower graphs, we also generated a graph that consists of 100,000 nodes. The results for that graph show that our method efficiently creates big graphs, exhibiting similar properties as real-world networks, which would likely take several weeks to generate with the old method.

We are convinced that the proposed method for the fast generation of directed social network graphs with reciprocal edges and high clustering to simulate the spread of epidemics or the circulation of information is a valuable addition to the list of network generation models. It has the flexibility to generate graphs of arbitrary size with similar topological features as real-world networks and it is highly scalable, which was demonstrated by generating a much bigger graph. 

\begin{acks}
This publication is part of the project "HPC and Big Data Technologies for Global Systems" (HiDALGO), which has received funding from the European Union’s Horizon 2020 research and innovation programme under grant agreement No. 824115. 
The Know-Center is funded within the Austrian COMET Program - Competence Centers for Excellent Technologies - under the auspices of the Austrian Federal Ministry for Climate Action, Environment, Energy, Mobility, Innovation and Technology, the Austrian Federal Ministry for Digital and Economic Affairs and by the State of Styria. COMET is managed by the Austrian Research Promotion Agency FFG.
The author thanks Bernhard C. Geiger (Know-Center GmbH) for his valuable feedback.\\
\end{acks}


\bibliographystyle{ACM-Reference-Format}
\bibliography{references}

\begin{thebibliography}{10}

\bibitem{Ahn2007}
Yong-Yeol Ahn, Seungyeop Han, Haewoon Kwak, Sue Moon, and Hawoong Jeong.
\newblock 2007.
\newblock Analysis of Topological Characteristics of Huge Online Social
  Networking Services.
\newblock In {\em Proc. Int. Conf. on World Wide Web (WWW)} (Banff, Alberta, Canada), ACM, New York, NY, USA, 835-844. https://doi.org/10.1145/1242572.1242685

\bibitem{Bansal2009}
Shweta Bansal, Shashank Khandelwal, and Lauren~Ancel Meyers.
\newblock 2009.
\newblock Exploring biological network structure with clustered random networks.
\newblock {\em BMC Bioinformatics} 10, 405 (2009). https://doi.org/10.1186/1471-2105-10-405

\bibitem{Barabasi1999}
Albert-Laszlo Barab\'asi and Reka Albert.
\newblock 1999.
\newblock Emergence of Scaling in Random Networks.
\newblock {\em Science} 286, 5439 (11 1999), 509--512. https://doi.org/10.1126/science.286.5439.509

\bibitem{Bender1978}
Edward~A Bender and E.Rodney Canfield.
\newblock 1978.
\newblock The asymptotic number of labeled graphs with given degree sequences.
\newblock {\em Journal of Combinatorial Theory, Series A} 24, 3 (1978), 296--307. https://doi.org/10.1016/0097-3165(78)90059-6

\bibitem{Bollobas1980}
Béla Bollobás.
\newblock 1980.
\newblock A Probabilistic Proof of an Asymptotic Formula for the Number of
  Labelled Regular Graphs.
\newblock {\em European Journal of Combinatorics} 1, 4 (1980), 311--316. https://doi.org/10.1016/S0195-6698(80)80030-8

\bibitem{Bonifati2020}
Angela Bonifati, Irena Holubov\'{a}, Arnau Prat-P\'{e}rez, and Sherif Sakr.
\newblock 2020.
\newblock Graph Generators: State of the Art and Open Challenges.
\newblock {\em ACM Comput. Surv.} 53, 2, Article 36 (March 2021), 30 pages. https://dl.acm.org/doi/10.1145/3379445

\bibitem{Britton2006}
Tom Britton, Maria Deijfen, and Anders Martin-Löf.
\newblock 2006.
\newblock Generating Simple Random Graphs with Prescribed Degree Distribution.
\newblock {\em Journal of Statistical Physics} 124 (October 2006), 1377--1397. https://doi.org/10.1007/s10955-006-9168-x

\bibitem{Chen2013}
Ningyuan Chen and Mariana Olvera-Cravioto.
\newblock 2013.
\newblock {Directed random graphs with given degree distributions}.
\newblock {\em Stochastic Systems} 3, 1 (2013), 147--186. https://doi.org/10.1214/12-SSY076

\bibitem{Chung2002}
Fan Chung and Linyuan Lu.
\newblock 2002.
\newblock Connected Components in Random Graphs with Given Expected Degree Sequences.
\newblock {\em Annals of Combinatorics} 6 (11 2002), 125--145.
\newblock https://doi.org/10.1007/PL00012580

\bibitem{Durak2013}
Nurcan Durak, Tamara~G. Kolda, Ali Pinar, and C.~Seshadhri.
\newblock 2013.
\newblock A scalable null model for directed graphs matching all degree
  distributions: In, out, and reciprocal.
\newblock In {\em Proc. IEEE Network Science Workshop (NSW)} 23--30. https://doi.org/10.1109/NSW.2013.6609190

\bibitem{Erdos1959}
Paul Erd\"{o}s and Alfred R\'{e}nyi.
\newblock 1959.
\newblock On Random Graphs I.
\newblock {\em Publicationes Mathematicae Debrecen} 6 (1959), 290--297.

\bibitem{Gilbert1959}
E.~N. Gilbert.
\newblock 1959.
\newblock Random Graphs.
\newblock {\em Annals of Mathematical Statistics} 30, 4 (1959), 1141--1144. https://doi.org/10.1214/aoms/1177706098

\bibitem{Het00}
Herbert~W. Hethcote.
\newblock 2000.
\newblock The Mathematics of Infectious Diseases.
\newblock {\em SIAM Review} 42, 4 (2000), 599--653. https://doi.org/10.1137/S0036144500371907

\bibitem{Kiss2017}
Istvan Kiss, Joel Miller, and Péter Simon.
\newblock 2017.
\newblock {\em Mathematics of Epidemics on Networks}, Volume~46.
\newblock Springer. https://doi.org/10.1007/978-3-319-50806-1

\bibitem{Kwak2010}
Haewoon Kwak, Changhyun Lee, Hosung Park, and Sue Moon.
\newblock 2010.
\newblock What is Twitter, a Social Network or a News Media?.
\newblock In {\em Proceedings of the 19th International Conference on World
  Wide Web} (Raleigh, North Carolina, USA), (WWW '10). Association for Computing Machinery page, New York, NY, USA, 591-600. https://doi.org/10.1145/1772690.1772751

\bibitem{Miller2019}
Joel~C. Miller and Tony Ting.
\newblock 2019.
\newblock Eon (Epidemics on Networks): a fast, flexible Python package for
  simulation, analytic approximation, and analysis of epidemics on networks.
\newblock {\em Journal of Open Source Software} 4, 44 (2019), 1731. https://doi.org/10.21105/joss.01731

\bibitem{Mislove2007}
Alan Mislove, Massimiliano Marcon, Krishna~P. Gummadi, Peter Druschel, and
  Bobby Bhattacharjee.
\newblock 2007.
\newblock Measurement and Analysis of Online Social Networks.
\newblock In {\em Proc. ACM SIGCOMM Conference on Internet Measurement (IMC)} (San Diego, California, USA). ACM, New York, NY, USA, 29-42. https://doi.org/10.1145/1298306.1298311

\bibitem{Myers2014}
Seth Myers, Aneesh Sharma, Pankaj Gupta, and Jimmy Lin.
\newblock 2014.
\newblock Information network or social network?: The structure of the twitter
  follow graph.
\newblock In {\em Proc. Int. Conf. on World Wide Web (WWW)} (Seoul, Korea). ACM, New York, NY, USA, 493--498. https://doi.org/10.1145/2567948.2576939
  
\bibitem{Newman2009}
Mark Newman.
\newblock 2009.
\newblock Random Graphs with Clustering.
\newblock {\em Physical review letters} 103 (07 2009), 058701. https://doi.org/10.1103/PhysRevLett.103.058701

\bibitem{Schweimer2022}
Christoph Schweimer, Christine Gfrerer, Florian Lugstein, David Pape, Jan~A.
  Velimsky, Robert Els\"{a}sser, and Bernhard~C. Geiger.
\newblock 2022.
\newblock Generating Simple Directed Social Network Graphs for Information
  Spreading.
\newblock In {\em Proceedings of the ACM Web Conference 2022} (Virtual Event, Lyon, France) (WWW '22). 
  Association for Computing Machinery, New York, NY, USA, 1475-1485.
\newblock https://doi.org/10.1145/3485447.3512194
  
\bibitem{Watts1998}
Duncan~J. Watts and Steven~H. Strogatz.
\newblock 1998.
\newblock Collective dynamics of ‘small-world’ networks.
\newblock {\em Nature} 393, 6684 (1998), 440--442. https://doi.org/10.1038/30918

\end{thebibliography}

\end{document}